\documentclass[journal, twoside, 10pt]{IEEEtran}

\ifCLASSINFOpdf
  
\else
  
\fi

\usepackage{amsmath}
\usepackage{hyperref}
\usepackage{graphicx}
\usepackage{bbm}
\usepackage{amssymb}
\usepackage[english]{babel}
\usepackage{lmodern}
\usepackage[T1]{fontenc}
\usepackage[utf8]{inputenc}
\usepackage{wasysym}
\usepackage{amssymb}
\usepackage{comment}
\usepackage{gensymb} 
\usepackage{array} 
\usepackage{url}

\usepackage[dvipsnames]{xcolor}

\usepackage{csquotes}

\usepackage[%
 bibencoding=utf8,
 backend=bibtex,
 bibstyle=ieee,
 citestyle=numeric-comp,
 defernumbers=true,
 sortcites=true,
 maxcitenames=1,
 mincitenames=1,
 maxbibnames=20,
 minbibnames=10,
]{biblatex}

\graphicspath{{Figures/}}

\newcommand{\rosso}{}   

\begin{document}

\title{\rosso{A Wearable Wireless Magnetic Eye-Tracker, \\ 
\textit{in-vitro} and \textit{in-vivo} tests}}



\author{
Giuseppe Bevilacqua,
Valerio Biancalana,
Mario Carucci,
Roberto Cecchi,
Piero Chessa,
Aniello Donniacuo,
Marco Mandalà,
Leonardo Stiaccini,
Francesca Viberti

\thanks{G. Bevilacqua, V. Biancalana, R. Cecchi and L. Stiaccini are with DSFTA, Siena University. Via Roma 56, 53100 Siena, Italy}
\thanks{M. Carucci, A. Donniacuo, M. Mandalà, and F. Viberti are with DSMCN,  UOC Otorinolaringoiatria, University of Siena,  Viale Bracci 16,  53100 Siena, (Italy)}
\thanks{P. Chessa is with Dept. of Physics "E.Fermi", University of Pisa, Largo Pontecorvo 3, 56127 Pisa, Italy}
\thanks{Manuscript received DATE1; revised DATE2.}
}

\markboth{Submitted 2023}{A Wearable Wireless Magnetic Eye-Tracker, \textit{in-vitro} and \textit{in-vivo} tests}

\maketitle

\begin{abstract}
A wireless, wearable magnetic eye tracker is described and characterized. The proposed instrumentation enables simultaneous evaluation of eye and head angular displacements. Such a system can be used to determine the absolute gaze direction as well as to analyze spontaneous eye re-orientation in response to stimuli consisting in head rotations. The latter feature has implications to analyze the vestibulo-ocular reflex and constitutes an interesting opportunity to develop medical (oto-neurological) diagnostics. Details of data analysis are reported together with some results obtained \textit{in-vivo} or with simple mechanical simulators that enable measurements under controlled  conditions. 
\end{abstract}


\IEEEpeerreviewmaketitle

\section{Introduction}
Tracking the movement of human eyes is relevant to a variety of research fields and  finds a range of practical applications. Among the scientific and health applications, the characterization of eye movement is of interest to different communities ranging from psychologists, neurologists, otolaryngologists, geriatricians, ophthalmologists. In fact, eye movement caused by external stimuli or activated to perform specific tasks provides useful information on various physiological and pathological conditions, related to attention, skills, vestibular response, degenerative diseases, etc. The breadth of this area of interest is related to the complexity of the eye dynamics (including fast / slow reorientations of small / large \rosso{angles} due to voluntary / unconscious actions), which is 
also related to the complex interconnection between peripheral  and central neurological systems involved in the control of eye movement. The reorientation of the eye is carried out with different types of movement which differ in terms of angular velocity and total displacement \cite{erkelens_vr_06}: namely \textit{pursuits} (slow movements along which the observed scene continues to be fixed) and \textit{saccades} (rapid movements that last a few milliseconds and cause very short, unperceived blindness intervals). 
Even during fixation, unperceived microsaccades occur without loss of vision. Duration and kinematics of fixation, latency, pursuits, saccades and microsaccades as well as their occurrence in response to assigned visual tasks or to other external stimuli provide information that can be used as diagnostic parameters  \cite{hutton_bc_08, ibbotson_con_11, termsarasab_jcmd_15, stuart_pm_16, stuart_ejn_18}.

The eye tracking  finds application also in robotics and related areas. 
In particular, it can be used to build human-machine interfaces, taking advantage from the speed and the spontaneous actuation of the eye orientation (e.g. in augmented and virtual reality) or to control prostheses and recover communication channels, since the eye motility persists in severely impaired patients.

Our brain elaborates the visually perceived information providing the impression that the whole accessible scene is continuously available with maximum detail and that it is recorded with some static or nearly-static device: we do not appreciate that the visual environment is constantly scanned by frequent saccadic movements in such way to project just a small angular region at once on the  fovea. In addition, the vestibular ocular reflex (VOR) stabilizes the visual area during the head movements, with opposite eye re-orientations. VOR helps to perceive static visual scene despite time dependent head-orientation, with a resulting subjective and illusive stillness of the scene. 
This biological system of visual stabilization hinders the study of actual ocular dynamics based on the perceived impression and requires the use of eye tracking techniques, the availability of which enables quantitative and scientific research on these otherwise elusive topics. 
It is worth mentioning that also smart alternative expedients enabling subjective evaluations of eye micromotion during fixation were proposed decades ago, based on after-image perception \cite{verheijen_oa_61}.

This research field has a long history \cite{wade_book_05}, and starting from the 18th century gained the attention of a number of scientists, who contributed both to characterize the eye dynamics and -- since the end of 19th century -- to build up refined instrumentation to quantitatively record the eye movements. Both ancient and modern techniques are historically presented in the first chapter of the above cited book, while a more detailed analysis of the currently used methodologies is reported in ref.\cite{eggert_no_07}. Among the latter, four complementary approaches emerge, namely electro-oculography, infrared reflection devices, video-oculography, and scleral search coil (SSC). 

These emerging methodologies are based on diverse kinds of measurements and are characterized by different levels of invasivity, accuracy, precision, speed, cost, need of pre-calibration etc., which cause them be more or less suited to be used in the various applications. The electro-oculography is based on measuring small voltages on electrodes applied in the eye proximity. It has the advantages of \rosso{low} invasivity and good time resolution and is not hindered by eye blinking, while it suffers from several noise sources and consequently offers limited precision and accuracy; it requires individual (patient-based) calibration. Infrared reflection devices and video-oculographs constitute other low-invasivity approaches (sometimes applied conjunctly \cite{guestrin_ieee_06, barsingerhorn_brm_18}) . The former use infrared (IR) sources and photodiodes to detect the IR light scattered by the cornea, the latter elaborate close-up images of the iris localizing the pupil center, hence both cannot work with closed eyes and produce artifacts at the eye blinkings.  Both kinds are commercially available for various uses -- including research, medicine, virtual reality -- and with different levels of spatial/time resolution (and cost). Both require individual calibration; the need of fast videocameras and cumbersome video processing make  video-oculographs with a high time resolution rather complex and expensive. 

The fourth methodology, the SSC, is based on the Faraday induction law: a pickup coil embedded in a scleral lens detects a multi-frequency alternating magnetic field. Field components are applied at different frequencies along two or three perpendicular axes, so that the search-coil orientation with respect to each single axis can be inferred from harmonic analysis of the induced electromotive force. The method has very good time and angular resolution, it is not hindered by eye blinking and does not require individual calibration (at least to infer the coil orientation: the need of a patient based calibration persists to derive the eye orientation from the coil one). On the other hand, the wired scleral lens makes the SSC particularly invasive. The Helmholtz coils that generate the homogeneous alternating field can be either fixed to the head (this makes the system wearable, but the bulky coils hinder head motion) or to a static frame surrounding the head: in both cases they constitute a severe constraint. The excellent performance associated with the high invasivity makes the SCC methodology essentially reserved to research applications, where it is considered the gold-standard technique. Unwired SSC detection based on double induction techniques has been proposed  \cite{bour_ieee_84, bremen_jnm_07}, which helps reduce the invasivity level but cannot solve the unwearability issue and the consequent intrusivity.

We have proposed an innovative tracking method based on determining position and orientation of a small magnet embedded in a scleral lens \cite{brevetto}. This methodology presents several advantages with respect to those mentioned above. The contact lens with the embedded magnet is unwired and hence much less invasive and intrusive than in the SSC case. The sensor frame is fixed to the head, but its light-weight makes the system wearable with negligible constraints to the head motion. Similarly to SSC (and differing from IR-reflection and video-oculography devices) it is not obstructed by eyelids and does not suffer from eye-blinking (apart for possible weak mechanical effects of the eyelids on the scleral lens). In analogy with SSC it does not require individual calibration concerning the magnet pose determination, while the need of a patient based calibration persists to convert the magnet pose into eye orientation \cite{biancalana_arxInstr_22}. The time resolution achieves 100~Sa/s in the current implementation, but can be extended to 200~Sa/s\cite{isentek8308} and even to 1~kSa/s with sensors recently commercialized \cite{isentek8308A}. The data elaboration does not require powerful machines (an ordinary personal computer enables real-time response) and the whole system is cost effective. The proposed instrumentation is at a prototypical stage, and feasible improvements may lead it to compete with other technologies in terms of angular and time resolution. Compared to the mentioned existing methodologies, it is placed at an intermediate level in terms of invasivity. Furthermore, the system provides simultaneous information about both eye and head orientation, allowing to determine the absolute gaze orientation and/or to analyze the eye motion in response to stimuli based on head movements. The latter feature is of particular interest to develop important otolaryngological diagnostics.

Tracking magnetically labeled devices has been proposed for other medical purposes not related to eye-tracking, possibly with relaxed requirements in terms of sampling rate, fast data elaboration and spatial resolution \cite{dinatali_ieee_13,gherardini_cmpb_21,meng_ieee_21, juce_sens_22, meng_ieee_22}. Wireless eye tracking based on magnetometric measurements for fast gesture estimation (not requiring high spatial or angular resolution) has been proposed as well \cite{tanwear_ieee_20}.

The instrumentation considered in this work has been previously described with emphasis on the hardware of and pre-calibration procedure \cite{biancalana_instrHW_21}, and on the data elaboration methodologies with their performance and limitations \cite{biancalana_instrSW_21} in view of generic applications. Having specialized the system to eye-tracking purposes, demonstrative tests were subsequently performed with a  simple eye-simulator \cite{bellizzi_rsi_22}, and preliminary \textit{in-vivo} were described in ref.\cite{biancalana_arxInstr_22}, which focused on the reconstruction of gaze trajectories. 

This paper focuses on the eye and head dynamics that can be investigated with this instrumentation. This subject has relevance in medical diagnostics and particularly to study physiological and pathological conditions of the eye movements,  and to estimate the VOR gain, with the subsequent implications in the evalutation of imbalance or in the early diagnosis of several neurological pathologies (such as Alzheimer's Disease \cite{opwonia_nr_22},  multiple sclerosis \cite{niestroy_clop_07} and Parkinson's disease  \cite{ba_fan_22}) and in the otoneurological field.
  
After having synthetically described in Sec.\ref{sec:setup}  the hardware and the software that produce tracking data, in Secs. \ref{sec:simulator} and \ref{sec:invivo} we present results obtained with mechanical head-eye simulators and preliminary recordings performed \textit{in vivo}, respectively. A conclusive Sec.\ref{sec:conclusion} summarizes the main results achieved and illustrates perspectives for further activity to be carried on.

\section{Setup}
\label{sec:setup}

\begin{figure}[ht]
   \centering
        \includegraphics [angle=0, width=  \columnwidth] {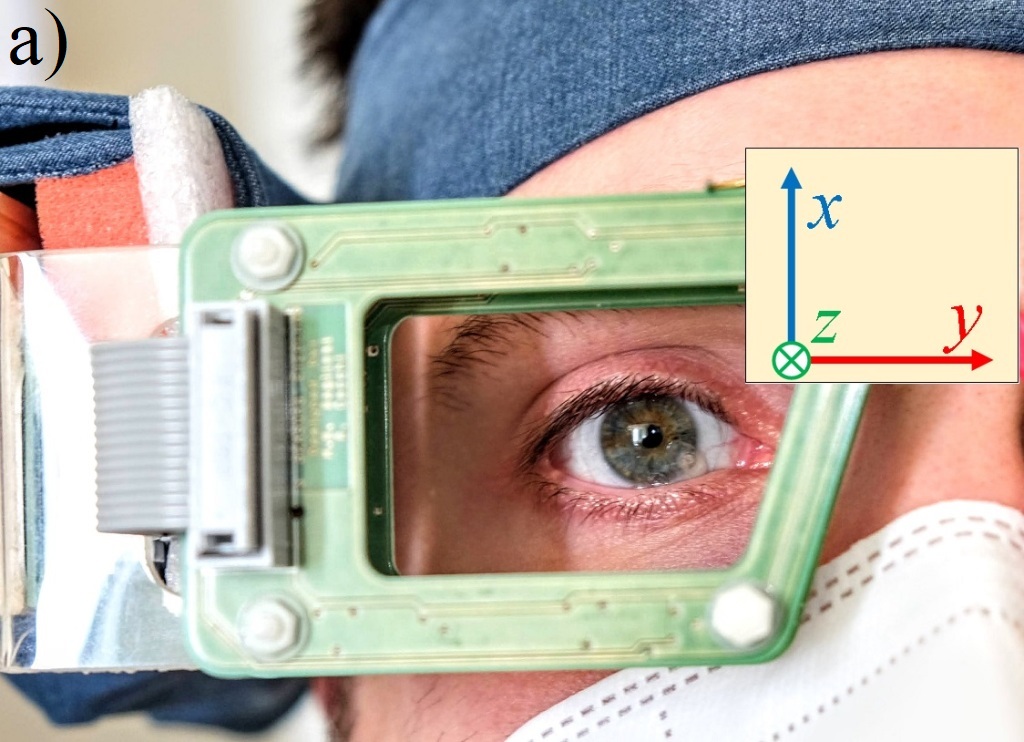}
        \includegraphics [angle=0, width=  \columnwidth] {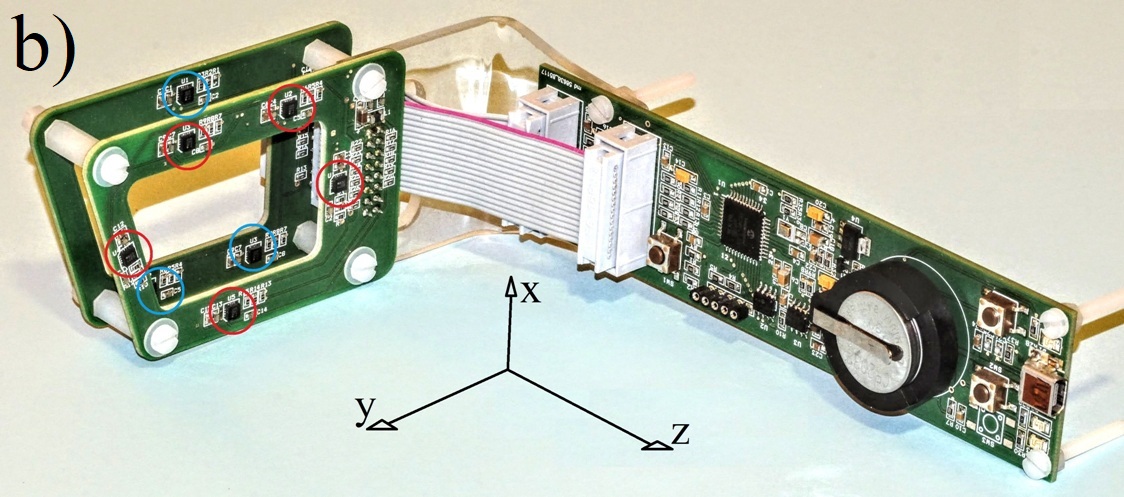}
        \caption{\rosso{a) The sensor array is rigidly fixed to the patient's head and tracks the movements of his right eye. b) It contains eight three-axial magnetoresistive sensors in two sets distributed on two parallel printed circuit boards (PCBs). Red and blue circles highlight the two sets. A third PCB hosts a microcontroller and other electronics that interface the sensors to a personal computer. The Cartesian axes with respect to the sensor array are shown in both the pictures. Nominally, when the head is erect and front oriented, $\hat z$ is back-directed, $\hat x$ is vertical and $\hat y$ is transverse-horizontal.} }
  \label{fig:sensorarray}
\end{figure}

\subsection{Sensors, hardware and firmware}
    The hardware (see Fig. \ref{fig:sensorarray}) has been extensively described in the ref.\cite{biancalana_instrHW_21} and additional details of its use as an eye tracker can be found in the refs., \cite{bellizzi_rsi_22, biancalana_arxInstr_22} 
     The sensor array has a goggle shape and contains eight three-axial magnetoresistive sensors, for a total of 24 data per measurement. \rosso{Raw data are affected by offsets and unequalized gains, thus a pre-calibration procedure is mandatory to convert them into magnetometric values (see Ref.\cite{biancalana_instrHW_21} for details). }
     A microcontroller card enables the communication of the sensors with a personal computer (PC). The latter is used to set the measurement parameters (rate, sensitivity etc.) and to \rosso{convert/store/elaborate the  readings}. Data elaboration and tracking analysis can be performed both on-line (with a nearly real-time response) and off-line. The data transfer to the PC uses a USB cable that also provides the system power supply, upgraded wireless versions with autonomous (battery) supply are feasible and are currently under design.
    
    The sensors used in the current prototype are Isentek 8308 \cite{isentek8308}, which have a maximum throughput rate of 200~Sa/s. However,  the rate tested so far is firmware limited to 100~Sa/s. It is worth noting that similar devices allowing acquisition rate up to 1~kSa/s are commercially available \cite{isentek8308A}. In this case the bottle-neck represented by the I$^2$C protocol is avoided by means of on-chip data buffering, and hence such a high rate is achievable only for  data bursts of limited size.
    
    The practical limitation set by the mentioned rates can be evaluated on the basis of typical eye angular speeds. The most demanding case is represented by saccades. Eye movements faster than a $40\degree$/s threshold are commonly cathegorized as saccades \cite{erkelens_vr_06}, but they may reach $700\degree$/s for large angular displacements in humans (see Sec.\ref{sec:VORdynamics}, and \cite{fuchs_jp_67} and refs. therein). Thus systems with accelerated data acquisition should be considered to analyze such fast saccades maintaining a $1\degree$ resolution. 
    In typical application (see Sec. \ref{sec:invivo}) saccades occur at about $200\degree$/s, so that the current 100~Sa/s acquisition rate guarantees a two-degree angular dynamic resolution. For comparison,the angular resolution in static conditions was estimated with in vitro measurements resulting in about 0.3\degree  as a precision and about 1\degree  as an accuracy \cite{biancalana_arxInstr_22}.

\subsection{From measurements to tracking parameters}
\label{subsec:bestfit}

Displacements and rotations of the target magnet are retrieved from a set of simultaneous field measurements performed in diverse, pre-assigned positions.  To this end, the field generated by the magnet is modeled in terms of a dipolar one, to which an (homogeneous) ambient field is superposed. The inverse problem of determining the dipole pose and the ambient field is faced by means of numeric tools, which were demonstrated to be adequately reliable. In particular we experimentally verified that the developed algorithms are sufficiently robust with respect to the initial guess and sufficiently fast to enable on-line data elaboration \cite{biancalana_instrSW_21}.

As far as the sensor array is  fixed to the head of the patient, the magnet pose (position and orientation) is determined in a co-ordinate system that moves rigidly with the head. Practical consequences are that the mechanical connection to the head requires particular care and that lighter arrays will help reduce systematic errors produced by head-sensor relative motion.

\subsection{From tracking parameters to eye- and head-pose}
\label{sec:gazeandhead}

In the hypothesis of a perfectly rigid array-head connection, the position $\vec r$ of the magnet, and the orientation $\hat m$ of its magnetic dipole $\vec m$ account for eye motion with respect to the head, while the retrieved $\vec B_\mathrm{geo}$ provide information about  the head orientation in the ambient field.

As discussed in ref.\cite{biancalana_arxInstr_22}, some limitations occur, which make impossible to strictly determine the  head orientation and the eye pose from the quantities retrieved by the tracker.
In particular, head rotations around the $\vec B_\mathrm{geo}$ orientation cannot be detected and rotations around an axis almost parallel to $\vec B_\mathrm{geo}$ would be barely detected. From a practical point of view, this means that, depending on the head rotations to be analyzed, an opportune orientation of the patient with respect to the ambient field could be necessary or advisable. 

Similarly, the orientation of $\hat m$ is related to the eye rotations with respect to the head, however it cannot directly provide the direction of the gaze $\hat e$, if the magnet is not oriented along the visual axis of the eye. It is not possible to establish a one-to-one relationship between  $\hat e$  and the measured $\hat m$. An intuitive proof of this limitation is realized if one considers the axial symmetry of the dipolar fields: eye rotations around the direction $\hat m$ would change the orientation of $\hat e$ while not causing any variation of $\vec m$ and therefore could not be detected.

Apart from the issues related to the mentioned \textit{blind} directions, provided that eye and head rotations are simultaneously retrieved from the tracking output, they can be combined or compared.
The combination may help infer the absolute gaze \cite{bellizzi_rsi_22}, while the comparison can be used to investigate eye rotations induced by head movements, and particular to characterize the VOR gain.

\subsection{VOR and other typical dynamics}
\label{sec:VORdynamics}

The VOR, one of the fastest reflexes in humans and vertebrates \cite{land_vr_19}, is fundamental to stabilize gaze during head and whole-body movements. The angular head displacement perceived by the vestibular system drives the oculomotor system to stabilize images on the retina within 10 msec by producing compensatory eye movements \cite{aw_jn_96}. VOR gain represents the amount of eye rotation relative to the amount of head rotation (in the correct plane) and it should be near unity (1.0), meaning that an equal and opposite eye rotation has been generated in response to a head rotation, which  ensures the perception of a stable scene  \cite{anson_fran_16}.
VOR analysis in term of latency, amplitude, velocity is a key factor in the diagnosis of neurotological disorders.
The VOR has a physiological latency as short as 8 ms \cite{hag_rvs_20}. An excessive latency (>20 ms) and/or an incorrect amount of the response constitute pathological conditions that may cause vertigo or dizziness, and are an evidence for vestibular \rosso{dysfunction} \cite{halmagyi_an_88}.

Clinical VOR evaluation is performed applying a sudden rotation (thrust) to the head of the patient (head impulse test, HIT) and evaluating compensatory  saccades.  It is worth mentioning that spontaneous eye micro-motions occur continuously, even without external stimuli  \cite{rolfs_vr_09}. They include relatively slow terms (tremors) -- with frequencies in the range of 70-100 Hz and amplitudes smaller than 0.1\degree (hence less than 1\degree/s typical angular velocities) -- and fast ones (microsaccades) that happen a couple of times per second with larger displacements (1\degree - 2\degree) and higher velocity, of the order of 10\degree/s. In addition, very slow drifts (< 0.5\degree/s) occur during the intersaccadic intervals.
The most demanding dynamics in terms of sampling rate is represented by large saccades. Sudden reorientations of the gaze may occur with large displacements (up to tens of degrees) lasting short time intervals: in that case, the angular velocity can ordinarily reach 100\degree/s - 500\degree/s range \cite{abrams_jep_89}.

\subsection{Rotations about an assigned axis}
\label{subsec:axisidentification}
In the current implementation, the magnetic dipole  embedded in the scleral lens is not parallel to the visual axis, but forms a small angle ($10\degree$-$30\degree$, typically) with it. As detailed in ref.\cite{biancalana_arxInstr_22}, under this condition the shape of a generic gaze trajectory does not coincide with that followed by the magnetic dipole: some assumptions (and consequent elaborations) are necessary to convert 2D dipole trajectories into gaze trajectories. This issue has a reduced importance when rotations about a single, assigned axis occur, i.e. with 1D trajectories.

A simple procedure to evaluate the matrix that describes the rotation  by an angle $\phi$ around an assigned direction $\hat u$ is based on the Rodrigues' formula \cite{rodrigues_jmpa_40}:

\begin{equation}
        \mathbf{R}=\mathbf{I}+\sin(\phi)\mathbf{K}+(1-\cos(\phi))\mathbf{K^2},
\end{equation}
being
\begin{equation}
       \mathbf{K}=
    \begin{bmatrix}
            0 & -u_z  & u_y\\
           u_z & 0 & -u_x\\
            -u_y & u_x & 0
    \end{bmatrix}
\end{equation}
the matrix such that, given a generic vector $\vec v$, $\mathbf{K} \vec v = \hat u \times \vec v$. 

In the HIT maneuvers applied for VOR gain evaluation, the rotation axis is expected to coincide with $\hat x$ (vertical axis, for \textit{pitch HIT}) or $\hat y$ (transverse horizontal for \textit{yaw HIT}), which leads to  simple $\mathbf{R}$ expressions, otherwise the actual rotation axis can be evaluated as it will discussed below. 

In any case, the same rotation matrix applies to both the magnetic dipole $\vec m$ and to the eye gaze direction $\hat e$, thus the rotation angle occurred between \rosso{two} subsequent measures of $\vec m$ can be evaluated and it is known that the corresponding \textbf{R} (i.e. the same rotation angle) applies also to $\hat e$. In particular, the rotation angle $\Delta \phi$ occurred between two subsequent measures $\vec m_1$ and $\vec m_2$ can be retrieved from the equation $\vec m_2 = \mathbf{R}\vec m_1$ or, more directly, from:
\begin{equation}
    \Delta \phi= \phi_{u,2}-\phi_{u,1}=\arcsin{ \frac{|\vec m_{2, \perp } \times \vec m_{1, \perp }|}{|m_{2, \perp }||m_{1, \perp }|}}
    \label{eq:arcsin}
\end{equation}
being
$\vec m_{i, \perp }= \vec m_i -(\vec m_i \cdot \hat u) \hat u $. 

The same analysis can be applied to two subsequent estimations of $\vec B_\mathrm{geo}$ to estimate the rotation $\Delta \theta$ of the head.

Optionally, if one expects that the actual rotation axis of the head may be at some angle from the nominal one, its direction can be evaluated from the whole set of measurements recorded during the maneuver. In particular, the actual rotation axis $\hat u$  can be identified on the basis of the $\vec B_\mathrm{geo}$ measurement set \{$B_i$\} ($i=1,\dots,N$) recorded during the maneuver. 
For instance, a least mean square fitting of the measured vectors on a plane \cite{pearson_js_901, schomaker_ac_59} provides an estimate of the rotation axis direction $\hat u$ as that perpendicular to the fitting plane. Of course, the larger is the rotation applied in the maneuver, the more accurate will be the estimation of $\hat u$.

The VOR gain estimation is usually based on comparisons of the head and eye rotation angle or of their time derivative.
It is worth noting that, considering the case of generic rotation, the VOR gain is not isotropic\cite{aw_jn_96, pogson_jn_19}. In particular, should the eye response require also torsional movements, that component of the eye rotation would be very depressed, which would result in a $\dot {\vec \phi}$ definitely not parallel to $\dot{\vec \theta}$.

As a consequence -- even in the simplifying hypotheses of a linear and non-delayed   response -- the VOR gain should be expressed tensorially: the VOR-gain defined on the basis of the ratio between angular velocities should be expressed as $\dot {\vec \phi}= \mathbf{G} \dot {\vec\theta}$, with $\mathbf{G}$  a non-trivial $3 \times 3$ matrix. Such a complete analysis is however beyond of the scopes of this work: we will proceed with a simplified treatment, assuming, like in most of the available literature \cite{aw_jn_96, anson_fran_16, hag_rvs_20},  that torsional rotations are avoided and that $\dot {\vec \phi}$ can be assumed to be parallel to $\dot { \vec \theta}$, i.e. that one rotation axis $\hat u$ is shared by eye and head.
The actual amount and influence of torsional rotations will be addressed in future studies after the implementation of an appropriate version of the magnetically labeled scleral lens. 
In fact, it is worth noting that the described tracking methodology, in conjunction with a transversely magnetized target (or with lenses hosting two displaced magnets) might enable an unprecedented sensitivity to torsional rotation \cite{biancalana_arxInstr_22}.

\section{Results with mechanical simulators}
\label{sec:simulator}

\subsection{Magnet position and absolute gaze}

We have performed several kinds of tests with simple mechanical simulators to evaluate the performance of the tracking system under controlled conditions.
Demonstrative tests for absolute gaze retrieval are presented in ref.\cite{bellizzi_rsi_22} and results that confirm submillimetric and sub-degree precision in the determination of the magnet position are reported in ref.\cite{biancalana_instrSW_21} and \cite{biancalana_arxInstr_22}, respectively. The former are obtained comparing the retrieved $\hat e$ with the projection of a laser beam, having applied the magnet to a spherical bearing that simulates a moving eye and holds laser pointer indicating the gaze direction. 
\rosso{The latter applies circular (or roto-translational, helical) trajectories to the magnet which allow to verify the accuracy with which the radius and the planarity (or the pitch) of the dipole trajectory are reconstructed. }
In all cases, the magnet displacements and the distance from the sensors were selected to be compatible with typical eye-tracking conditions.

\subsection{Magnet orientation and VOR-gain evaluation}
\begin{figure}
\centering
 
        \includegraphics [angle=0, width= \columnwidth] {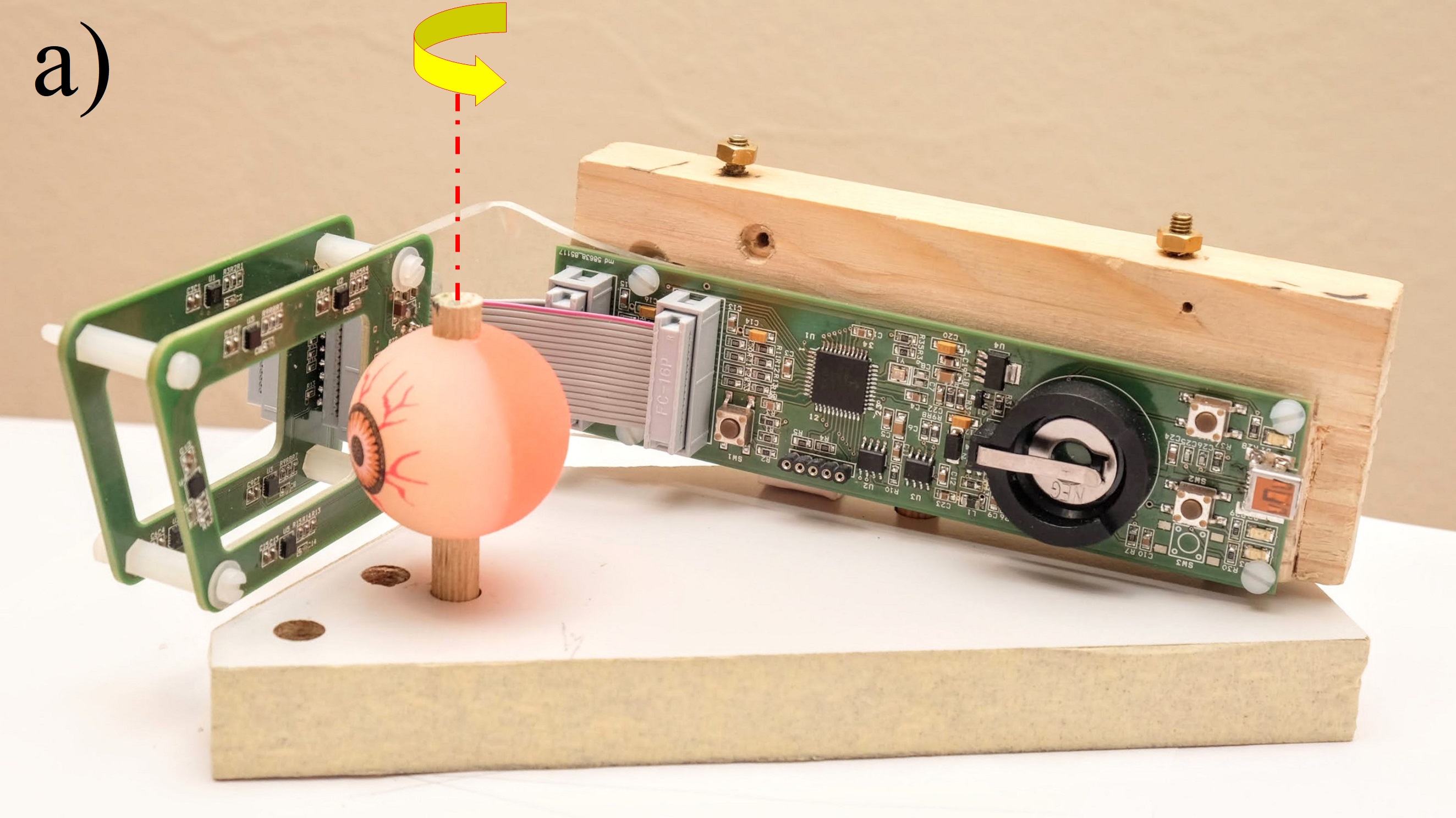}
        \includegraphics [angle=0, width= \columnwidth] {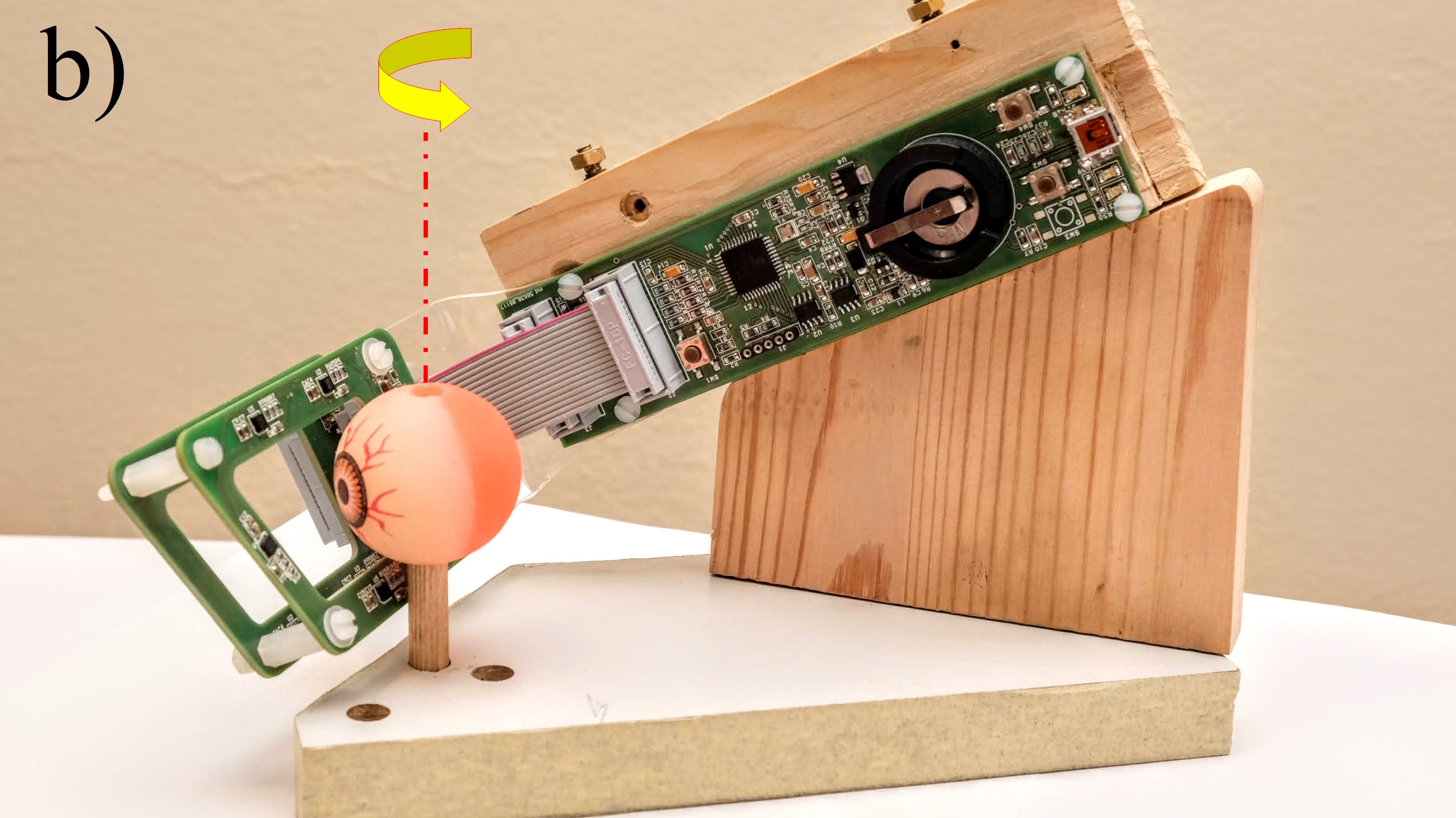}
      \caption{A simple simulator enables the assessment of the system performance in the identification of the rotation axis and in the evaluation of the VOR gain. The sensor array is rotated around a vertical axis (red line) that (a) coincides  with one of the axis of its reference frame or (b) is generically oriented. This simulates the head rotation in an ideal or non-ideal HIT maneuver, respectively. A standing magnet fixed at 12~mm from the rotation axis simulates the case of an eye motion that perfectly compensates the head rotation. \rosso{The magnet is glued on the eye-phantom, in near front direction, on the hidden side in these pictures} }
      \label{fig:VORideale}
\end{figure}

A third simulation  setup makes possible to assess the performance of the system and the reliability of the data analysis procedures, in measurements that simulate VOR gain analyses. To this  purpose, the sensor array is rotated around a vertical axis and the magnet is located 12~mm away from that axis. The axis can be either parallel to the $\hat x$ direction (Fig.\ref{fig:VORideale} a) or generically oriented with respect to the sensor co-ordinate frame (Fig.\ref{fig:VORideale} b).

This simple device is hand actuated and a goniometric scale makes it possible to control the applied angular displacements, while the angular speed is established by the operator, but is not directly measured. This arrangement is used to assess the reliability of the rotation axis identification procedure (Sec.\ref{subsec:axisidentification}), and to infer accuracy and precision of VOR gain estimations.

\begin{figure}
\centering
        \includegraphics [angle=0, width=\columnwidth]{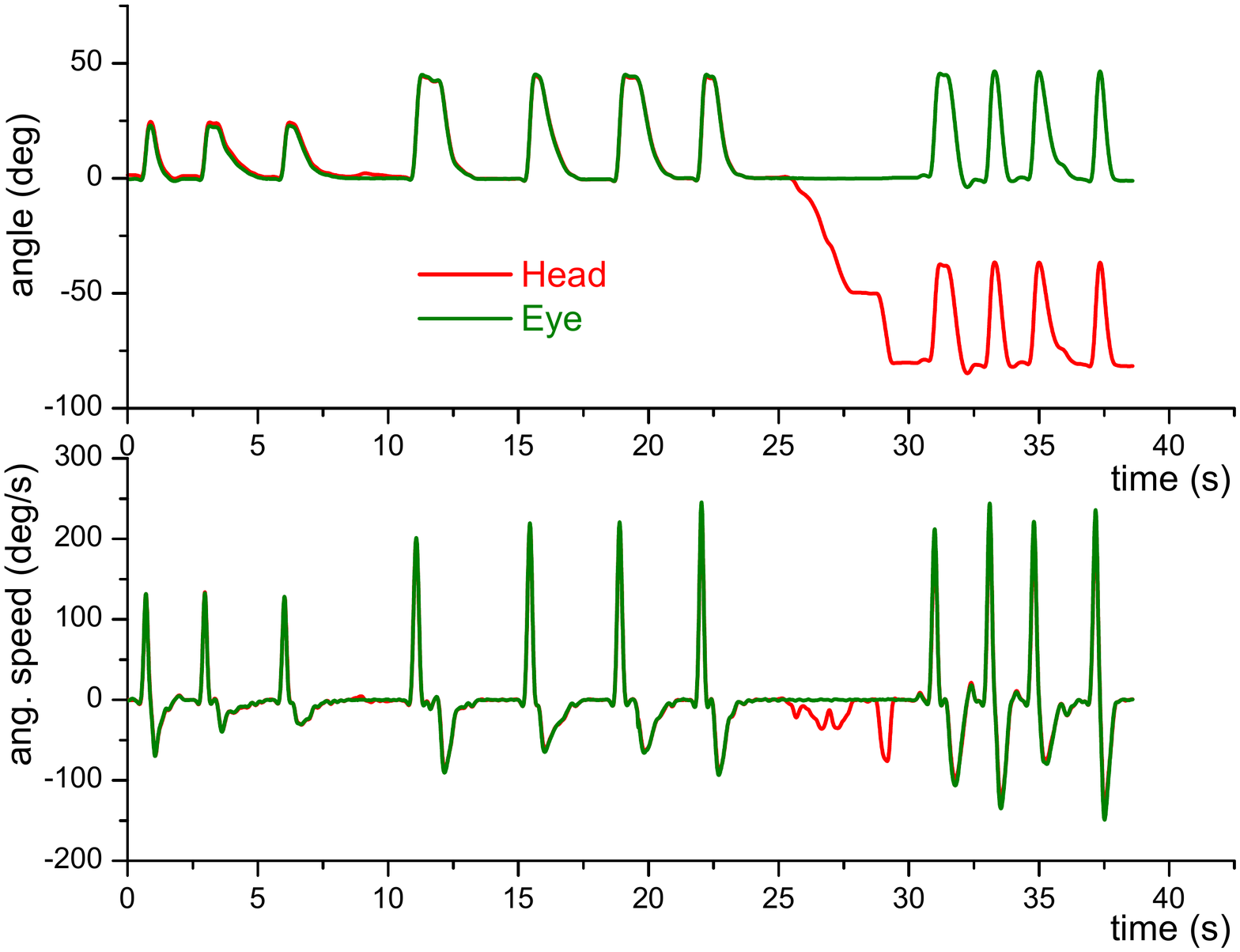}
  \caption{Simulation of VOR-gain measurement. The measurement consists in three $22.5\degree$ maneuvers; four $45\degree$; a rotation of the whole system, and four additional $45\degree$ maneuvers. The plots show the measured angles and angular speeds of $\vec m$ (eye, green) and $\vec B_\mathrm{geo}$ (head, red), respectively. The angles are evaluated around an inclined rotation axis (Fig.\ref{fig:VORideale}b),  previously identified on the basis of the $\vec B_\mathrm{geo}$ values measured at the 5$^\mathrm{th}$ maneuver (see Sec.\ref{subsec:axisidentification}).
  } \label{fig:vorlegno}
\end{figure}

The rotation of the array around the fixed axis reproduces the head rotation, and the standing magnet orientation simulates the case of an ideal (unitary gain) VOR. When operating in the configuration (a), head and eye rotation can be evaluated around the $\hat x$ direction, while in the configuration (b) it is strictly necessary to analyze the data after having identified the rotation axis $\hat u$ following the procedure described in Sec.\ref{subsec:axisidentification}. 
This latter is the case of the results shown in Fig.\ref{fig:vorlegno}, which shows HIT simulations performed with $22.5\degree$ and $45\degree$ rotations around a generically oriented axis. The last four maneuvers have been performed after having rotated the whole system, which demonstrates the robustness of the apparatus with respect to the orientation of the environmental field.
Under these favorable, controlled conditions, both the gain estimated on the basis of the angular displacement and on the peak velocity result unitary within a 1 \% uncertainty. 

The literature reports several definitions of VOR gain\cite{pogson_jn_19}. They are  based on comparison of angular displacements \cite{mcgarvie_fr_15} or peak angular speeds \cite{aw_jn_96, hag_rvs_20}. Different choices are made to define the beginning and the end of the HIT maneuver \cite{cleworth_jvr_17} and, when speeds are compared, they are evaluated at their peak values or at a given delay with respect to instant of the peak value. 

Velocity evaluation based on numerical differentiation requires preliminary denoising and/or data filtering, whose settings can  non-negligibly affect all the mentioned values and eventually the VOR gain estimation. 
\rosso{In addition, a physiological VOR must guarantee a fast recovery of the gaze orientation. This suggests that a VOR evaluation based on the comparison of head and eye angular displacements within a short time interval surrounding the HIT \cite{mcgarvie_fr_15} is a favorite choice.}

\section{Results \textit{in vivo}}
\label{sec:invivo}

The data reported in this section have been recorded \textit{in vivo} with a help of a healthy volounteer (one of the authors), wearing a scleral lens with a 2~mm diameter 0.5~mm thickness axially polarized Nd-magnet embedded. The acquisition rate was set to the maximum value  (100 Sa/s).

\subsection {Gaze trajectory}
\begin{figure}
\centering
        \includegraphics [angle=0, width=0.49\columnwidth] {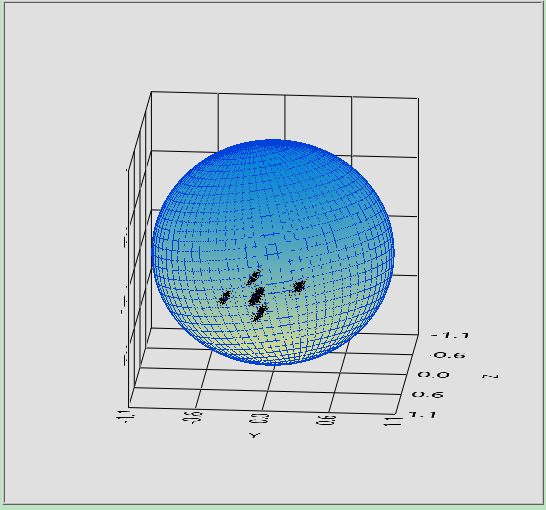}
        \includegraphics [angle=0, width=0.49\columnwidth] {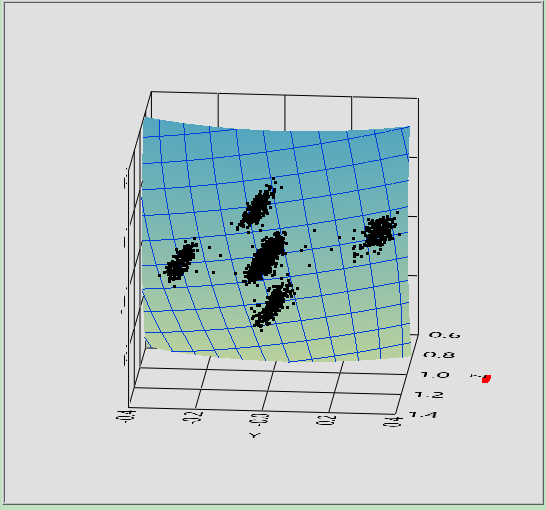}
        \includegraphics [angle=0, width=0.49\columnwidth] {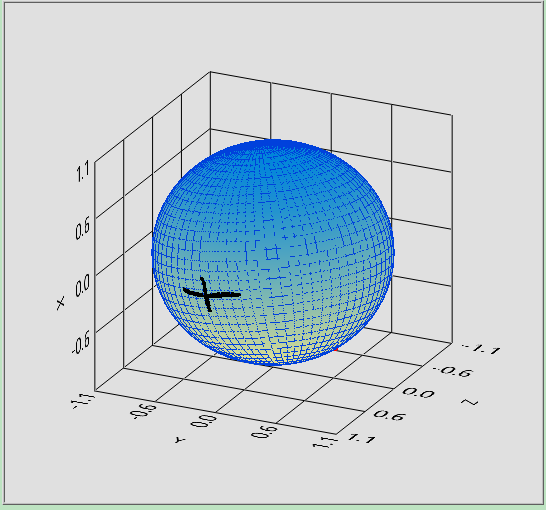}
        \includegraphics [angle=0, width=0.49\columnwidth] {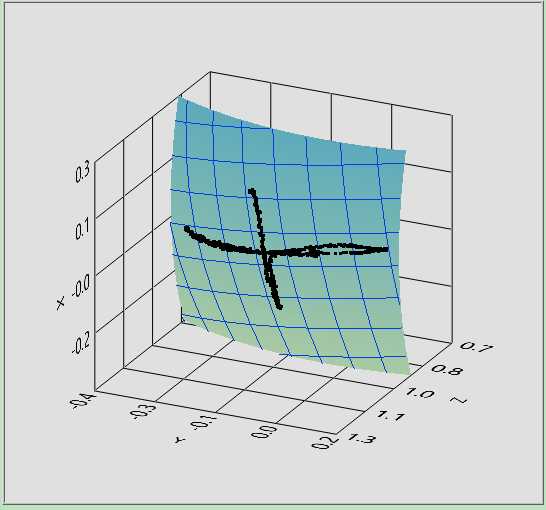}
  \caption{Retrieved gaze direction from measurements obtained with standing head when the subject is requested to follow a point that moves suddenly (saccades, upper panels) or smoothly (pursuits, lower panels) along a cross-shaped trajectory. The grid squares measure $5\degree \times 5\degree$ both in the wide-view (left panels) and close-up (right panels) representations.
  } \label{fig:mariosaccpurs}
\end{figure}

Examples of gaze reconstructions are represented in Fig.\ref{fig:mariosaccpurs}. The subject was requested to fix a point target moving on a monitor while keeping the head standing. The point followed a cross-shaped trajectory (center-right-center-left-center-down-center-up-center) either in jumps or with continuity. The subject was about 1~m from the screen and the cross size was $2 \times 30$~cm in width and $2 \times 18$~cm in height. 

\subsection{Saccades and pursuits}
\begin{figure}
\centering
        \includegraphics [angle=0, width=\columnwidth] {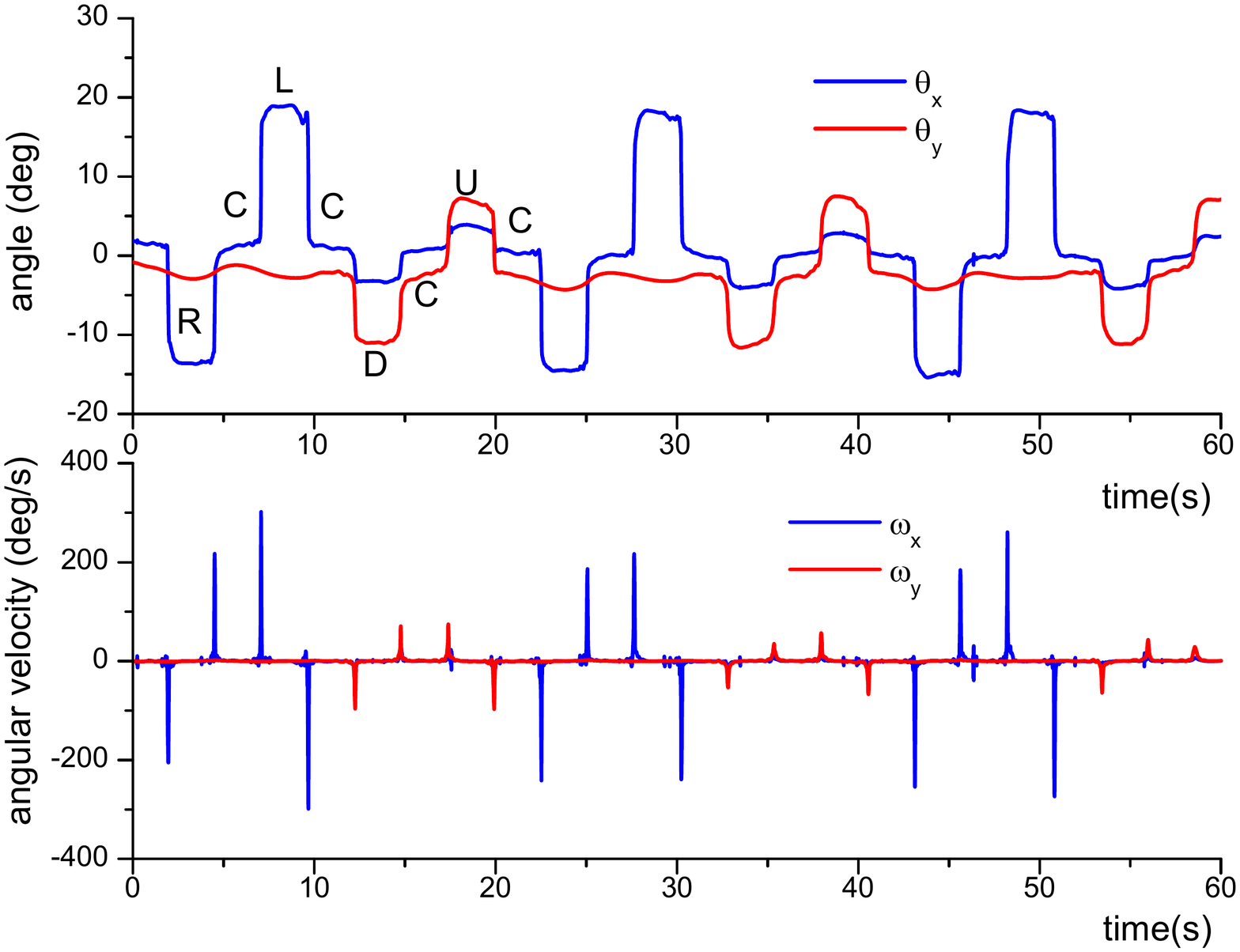}
  \caption{The same data represented in the upper panels of Fig.\ref{fig:mariosaccpurs} are here plotted as a function of time, evaluating rotation angles (and their first time-derivative) analyzing rotations around a vertical axis (blue traces) or a horizontal one (red traces), respectively. Capital letters identify the gaze orientation (R=right, C=center, L=left, D=down, U=up)
  } \label{fig:SACCvsT}
\end{figure}

\begin{figure}
\centering
        \includegraphics [angle=0, width=\columnwidth] {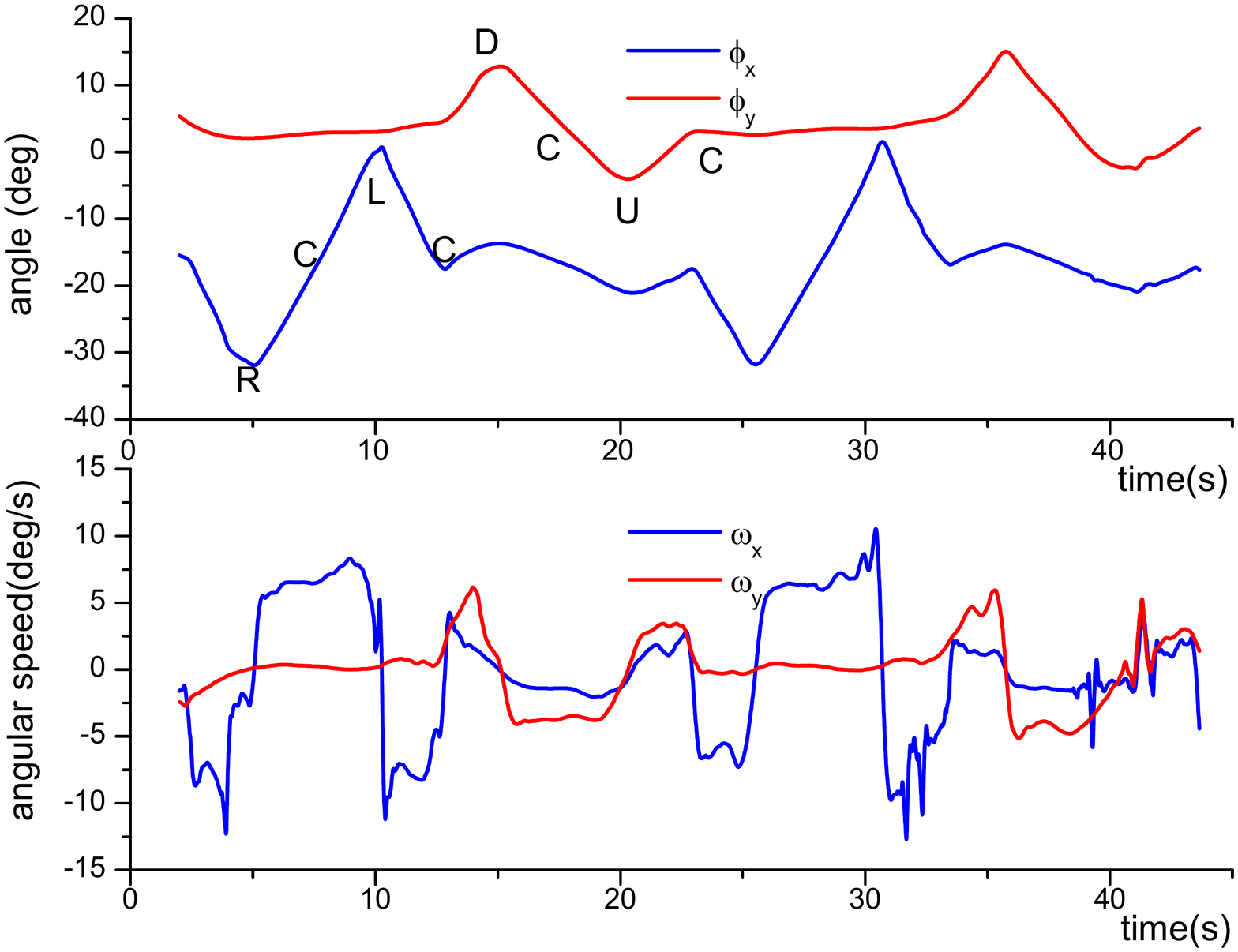}
  \caption{ The same data represented in the lower panels of Fig.\ref{fig:mariosaccpurs} are here plotted as a function of time, evaluating rotation angles (and their first time-derivative) analyzing rotations  around a vertical axis (blue traces) or the horizontal one (red traces), respectively. 
  } \label{fig:PURSvsT}
\end{figure}

The same data represented in the 2D plots of Fig.\ref{fig:mariosaccpurs} are used, after the application of wavelet-based denoising techniques \cite{donoho_ieee_05}, to produce the graphs shown in Figs.\ref{fig:SACCvsT} and \ref{fig:PURSvsT}, where angular displacements and angular velocities are plotted versus time, respectively. The angles are retrieved according to the Eq.\ref{eq:arcsin}, assuming $\hat u= \hat x$ or $\hat u= \hat y$, respectively.

The estimations on the dynamics of both pursuits and saccades are in quantitative agreement with the expected tasks to be performed by the subject. In particular the angular displacements represented in Figs.\ref{fig:SACCvsT} and \ref{fig:PURSvsT} and the slopes in Fig.\ref{fig:PURSvsT} substantially match the amounts set by the trajectory followed by the moving point and by the subject-monitor distance. 

As said, these estimates are performed separately, analyzing either  the eye rotations around  a vertical axis (blue traces) or around a horizontal one (red traces). 
In both cases a weak contamination appears between horizontal and vertical rotations. In particular, up-down movements are partially detected as $\phi_x$ rotations. This imperfection is likely due to a residual system misalignment, which can be also appreciated in the inclination of the estimated polar axis (as shown in  Fig.\ref{fig:mariosaccpurs}). 

\subsection {VOR gain measurement}
\begin{figure}
\centering
        \includegraphics [angle=0, width=\columnwidth] {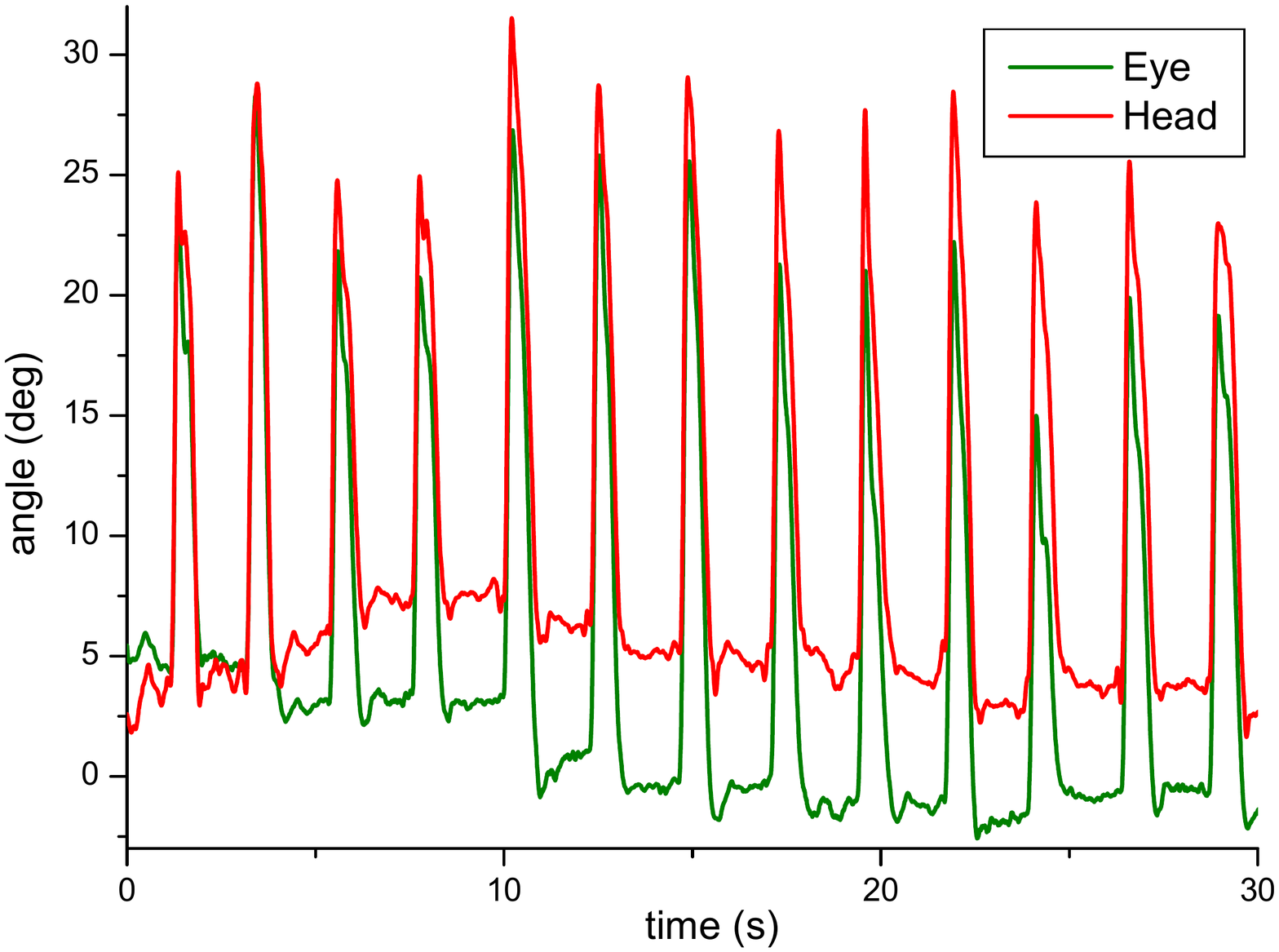}
  \caption{Results of \textit{in vivo} HIT maneuvers for VOR gain evaluation. 
  } \label{fig:vorMario}
\end{figure}

The system -- worn as shown in Fig.\ref{fig:sensorarray} -- was tested in a real set of measurements for estimating the VOR gain. Fig.\ref{fig:vorMario} shows angular displacement of head and eye retrieved in a set of HIT maneuvers consisting in fast rotations (about $20\degree$) of the head around a nearly vertical ($\approx \hat x$) axis, which induce opposite (compensating) saccades driven by VOR. The gain is evaluated as the ratio between eye and head angular displacements, which are estimated by means of \rosso{Eq.}\ref{eq:arcsin}. In this set of 13 samples is $G_\mathrm{VOR-angle-left}=1.01 \pm 0.07$ (mean value $\pm$ standard deviation). For comparison, the VOR-gain estimated with the same data on the basis of peak-velocity ratios is  $G_\mathrm{VOR-speed-left}=1.28 \pm 0.12$. All the maneuvers considered in this test consist in sudden rotations of the head towards the left side of the patient.
Incidentally, maneuvers performed in the opposite sense resulted in unexpectedly larger (and wrong) $G_\mathrm{VOR-angle-right}$ and $G_\mathrm{VOR-speed-right}$. Such overestimation was most likely caused by a slippage caused by inadequate mechanical array-head connection: a practical detail that requires additional care.

\section{Perspectives and Conclusion}
\label{sec:conclusion}

\subsection {Perspectives}

Diverse tests of the performance of a prototypical wireless wearable eye-tracker have been described. These tests confirm interesting potentials of the developed instrumentation and help identify aspects to be improved in next implementations. 

The data-acquisition rate constitutes a limitation when investigating saccadic movements that cause large angular velocities. Minor adjustments in electronics and firmware are expected to enable a twice higher rate, while new types of sensors should enable, at least for data bursts, a 10 times increase on the basis of the same protocol and architecture.

The current prototype was designed with the sensors distributed along three directions thanks to a dual PCB structure (two layers of sensors). However, we have verified that the tracking algorithm maintains a good reliability even when only data from sensors laying on one PCB are used. This observation suggests to use a single-PCB structure, which will result in a lighter device and in a wider free visual angle.

In VOR maneuvers, a minimized inertia of the device would reduce artifacts due to array slippage with respect to the head. In next designs, soft connection between sensor and microcontroller PCBs will make necessary to guarantee a perfectly solid connection only for the (unburdened) PCB that hosts the sensors.

\rosso{The wireless nature of the current prototype is limited to the eye-sensor connection, which is crucial to reduce the invasivity with respect to SSC approaches. We plan to use also wireless data communication between microcontroller and PC, which will make the system fully unwired, with further advantages in terms of non-intrusivity.}

As discussed in Sec.\ref{sec:gazeandhead}  the ambient field may result disadvantageously oriented and cause a reduced accuracy in the tracking of the head movements. A gyroscopic sensor could be included in the device to generate complementary, field-independent data providing an additional/alternative measure of head rotations.

An added value of the \rosso{proposed} methodology resides in the wearability and in the low intrusivity. These positive characteristics can be further enhanced equipping the device with an autonomous power supply and a wireless data-transfer unit.

A practical aspect that needs care and perfecting concerns the production of scleral lenses with embedded magnet. So far, we provided ourselves with rigid lenses where a small hole is drilled to host a glued magnet. Two drawbacks consist in the need for individual (thus expensive) tailoring of such lenses, and in possible damaging of the magnet coating. We verified that the latter makes the magnet chemically unstable in the lens solution. Other methodologies for magnet insertion and  the use of soft (non tailored) lenses are worth of being investigated, as also reported recently for a similar application \cite{tanwear_ieee_20}. The solidity of eye-lens and head-sensor connections is a key feature that requires particular care and an accurate validation.

The embedding procedure could also be improved with efforts to align the magnet dipole with the gaze. Having $\vec m$ structurally parallel  to $\hat e$ would facilitate and simplify the interpretation of the tracking data, as discussed in ref.\cite{biancalana_arxInstr_22}.

At the other extreme using diametrically magnetized magnets to make $\vec m  \perp \hat e$ would pave the way to an attractive perspective, because it would make the system highly responsive to torsional movements. This feature is of particular interest, because such a response is hard (if not impossible) to be achieved  with the competing technologies \cite{ong_jnm_10, jin_inf_20}. To the same end, setups with two or more magnets separately embedded in the lens can be considered.

Efforts are planned to provide further and more detailed characterization of the system performance. The implementation of digitally controlled eye and head simulators would enable  other quantitative assessments, particularly for angular and time resolution.

\subsection {Conclusion}
We are \rosso{developing} an innovative eye-tracker based on non-inductive magnetometric measurements performed in multiple, preassigned positions with the aid of an array of magnetoresistive sensors. 
We have tested the device to record pursuits and saccades and to evaluate VOR gain in HIT maneuvers. Both with mechanical simulators and with preliminary \textit{in vivo} measurements.

We have analyzed advantages and disadvantages of the proposed instrumentation and we have pointed out several positive features (concerning precision, speed,  robustness to artifacts, and potentiality for an enhanced response to torsional movements) and negative ones (invasivity, intrusivity) with an analysis aimed to assess the achieved compromise, also in comparison with those achievable by concurrent eye-tracking technologies.

\vspace{6pt} 

\section*{Acknowledgment}
The authors are pleased to thank Y.Dancheva, who carefully read the manuscript suggesting important improvements.

\printbibliography[heading=none, notkeyword=OWN]



\end{document}